\documentclass[twocolumn,runningheads,natbib]{svjour2}
\smartqed  
\usepackage{graphicx}
\journalname{Astrophysics and Space Science}
\begin{document}

\title{Neutron star masses: dwarfs, giants and neighbors 
}


\author{Sergei  Popov         \and
        David Blaschke \and
        Hovik Grigorian \and
        Mikhail  Prokhorov 
}


\institute{S.B Popov \at
              Sternberg Astronomical Institute \\
              119992 Russia, Moscow, \\
              Universitetski pr. 13\\
              Tel.: +7-495-9395006\\
              Fax: +7-495-9328841\\
              \email{polar@sai.msu.ru}           
           \and
           D. Blaschke \at
Gesellschaft f\"ur Schwerionenforschung mbH (GSI),
D-64291 Darmstadt, Germany, and\\
Bogoliubov Laboratory for Theoretical Physics, JINR Dubna, 141980 Dubna,
Russia\\
\email{Blaschke@theory.gsi.de}           
           \and
           H. Grigorian \at
Institut f\"ur Physik, Universit\"at Rostock\\
D-18051 Rostock, Germany, and\\
Department of Physics, Yerevan State University, 375049 Yerevan, Armenia\\
             \email{hovik.grigorian@uni-rostock.de}           
           \and
           M.E. Prokhorov \at
            Sternberg Astronomical Institute \\
              119992 Russia, Moscow, \\
              Universitetski pr. 13\\
              \email{mike@sai.msu.ru}           
}

\date{Received: date / Accepted: date}

\maketitle

\begin{abstract}

We discuss  three topics related to the neutron star (NS) mass spectrum.
At first we discuss the possibility to form low-mass
($ M \stackrel{<}{\sim}   1\,M_{\odot}$)
and suggest this is possible only due to fragmentation
of rapidly rotating proto-NSs.
Such low-mass NSs should have very high spatial velocities
which could allow identification. 
A critical assessment of this scenario is given.
Secondly, we discuss mass growth due to accretion for NSs in close binary systems.
With the help of numerical population synthesis calculations we derive the
mass spectrum of massive ($M > 1.8\, M_{\odot}$) NSs.
Finally, we discuss the role of the mass spectrum in population studies of young
cooling NSs. We formulate a kind of {\it mass constraint} which can be
helpful, in our opinion, in discussing different competive models of the
thermal evolution of NSs.

\end{abstract}

\section{Introduction}
\label{intro}

 Mass is one of the key parameters
for neutron star (NS) physics and astrophysics.
It can be measured with high precision in binary radio pulsar systems.
Until very recently all estimates were obtained in the very narrow region of
1.35-1.45~$M_{\odot}$ (\citealt{tc1999}).
These values lie very close to the
Chandrasekhar limit for white dwarfs.
Thus, $M=1.4\, M_{\odot}$ was considered to be the
standard value of the NS mass.
Recently, the range widened towards lower masses
after the discovery of the double pulsar J0737-3039
(\citealt{b2003}).
One of the NSs in this system has $M=1.25\, M_{\odot}$
(\citealt{l2004}).
Several other examples are known (see Sec.~4).

The mass range is extended also towards higher masses, although
these results are less certain.
There is  one NS in a binary radio pulsar system with  mass
significantly higher than the canonical value $1.4\, M_{\odot}$.
It is the pulsar J0751+1807  with the mass
$2.1^{+0.4}_{-0.5}$ (95\% confidence level) (\citealt{ns2004}).
A recent reanalysis of RXTE data suggests that in low-mass X-ray
binaries there are several candidates for neutron stars with masses
between 1.9  and 2.1 $M_\odot$ (\citealt{bom2005}) as, e.g., 4U 1636-536.
The small number of massive radio pulsars, however,
can be a result of selection effects.

Cooling curves of NSs are strongly dependent on the star mass,
the mass spectrum is one of the most important ingredients of the population
synthesis of these sources and,
unfortunately, one of the less known.
Still, theoretical considerations \citep{whw2002}
can shed light on the general properties
of the mass spectrum of NSs. This properties have to be taken into account
when confronting theoretical models of the thermal evolution of NSs with
observational data.

In this paper we discuss three issues: the formation of low-mass NSs,
the mass growth of NSs in binaries, and the role of NS masses in the population
synthesis of young cooling objects.


\section{Formation of low-mass neutron stars}
\label{sec:1}

The results presented in this section are based on the research note
\citep{popov04}.

In many models of thermal evolution
of compact objects (neutron stars -- NS,
hybrid stars -- HyS, strange stars, -- SS) low-mass sources with
$0.8\, M_{\odot} <M<1\, M_{\odot}$
\footnote{Here and below speaking about compact objects we mean the
gravitational mass.}
remain hot for a relatively long time (about few million years)
\footnote{Objects with even lower mass, $\sim 0.5\, M_{\odot}$,
 also are relatively hot \citep{bgv2004}.}.
During all that time they remain hotter
than more massive stars  (\citealt{bgv2004} and references therein).
In that sense they are promising candidates to be
observed as {\it coolers}, and their detection is of great interest for the
physics of dense matter \citep{chp2003}.
However, in most of models of NS formation \citep{whw2002, fk2001, tww1996}
no objects with $M<1$~--~1.2~$M_{\odot}$ are formed.
It is likely that just because masses of
stellar cores are always heavier than 1.2~$M_{\odot}$ even for the solar
metallicity (and heavier for lower metallicity, see \citealt{whw2002}).

In our opinion
the only way to form a low-mass object from a (relatively) high-mass core
is fragmentation (see however a discussion in \citealt{x2004},
where the author
discusses the formation of low-mass SSs from white dwarfs via accretion induced
collapse).

The fragmentation of a rapidly rotating proto-NS due to
dynamical instabilities as part of a two-stage supernova
(SN) explosion mechanism was suggested by \cite{betal1988}, later the
mechanism was developed by \cite{in1992}.
This mechanism can explain several
particular features of SN explosions (for example the delay in the
neutrino signal from SN1987A).
In this scenario a compact object inevitably obtains a high kick velocity.
Recently, the mechanism
was studied in some details by \cite{cw2002} in application to kick phenomenon.
Because of high temperature the exploding (lighter) companion
can be significantly more massive than the minimum mass for cold stars
($\sim 0.1\, M_{\odot}$), up to $\sim 0.7 \,M_{\odot}$.
In that case from the initial object of, say,
1.2~--~1.3 solar masses due to fragmentation and explosion of the lighter part,
we can finally obtain a high-velocity NS with a
mass of about 0.8-1~$M_{\odot}$ or
lower if the mass of the initial object was smaller.

 Even lower masses can appear if in the fragmentation three bodies are
formed. In such a case the lightest or an
intermediate mass fragment can be dynamically ejected
from the system (again with significant velocity about thousands km~s$^{-1}$).
Such ejected compact objects can have masses about 0.2--0.5 solar masses.
In the remaining pair the lighter one can start to accrete onto the second
companion because of the orbit shrinking due to gravitational waves emission,
and after reaching the minimum mass ($\sim 0.1\, M_{\odot}$)
it explodes. So, the remaining compact object would also have relatively
low mass ($\sim 1\, M_{\odot}$) and high velocity.

 Objects formed after fragmentation have particular predictable properties:
high spatial velocity, high surface temperature, velocity vector nearly
perpendicular to the spin axis (since in this mechanism the kick is always obtained
in the orbital plane which coincides with the equator of the initial proto-NS).

 Due to high kicks low-mass compact objects are not expected to appear in
binaries (at least they should be rarely found in binary systems).
 To find a low-mass compact object one has to search for a
hot, young, high velocity NS.

 It is reasonable to expect that mass and kick velocity are
anti-correlated, as far as a higher mass of the remaining object
corresponds to the lighter exploded
component, which means to a wider orbit, and  to lower orbital velocity
of the remaining more massive component. Also higher kicks lead to smaller
fall-back (Colpi, Wasserman 2002).

 To reach fragmentation conditions (the dynamical  instability)
it is necessary that the progenitor core
is rapidly rotating. Rotation of isolated progenitors and its influence on
properties of newborn NSs was studied in several papers (see, for example,
Heger et al. 2003 and references therein).
To obtain a rapidly rotating compact object it is necessary to avoid
spin-down influence of the magnetic field, so probably compact objects born
after fragmentation should be low magnetized.
It means that low-mass neutron
stars are not expected to be normal radio pulsars. Because of the same
reason they are not expected to show any kind of
magnetar activity.

We have to note, that the mechanism of SN explosion
suggested by \cite{betal1988} and \cite{i1992}
has its internal problems.
There is the possibility, that a rapidly rotating proto-NS can just
loose part of its mass in the form of an  outflow in the equatorial plane.
In that case two spiral arms appear, no
second (or third) component is formed, and the kick can be relatively small.
The fraction of lost mass is very small, about 4\% (Houser et al. 1994),
so that the final mass cannot be much lower than the initial one.
If the fragmentation in the process of NS
formation never happens in nature, then, in our opinion, it is very
improbable, that low-mass compact objects can exist.
The discovery of a high velocity low-mass NS, HyS
or SS will be a strong argument in favour of the  mechanism.

To conclude: the fragmentation of a proto-NS can be a unique mechanism of
the formation of low-mass compact objects, which are expected to have
several peculiar characteristics that can help to distinguish them among
possible candidates. 
However, the realization of this mechanism in Nature is not very promising.


\section{Massive neutron stars in binaries}

In this section we present our recent calculations of the mass growth of NSs
in close binary systems \citep{pp05}.

Observationally, high masses of NSs are mainly
supported by data on X-ray binaries,
where recently new results have been extracted from a
reanalysis of RXTE data which has suggested that sharp and reproduceable changes
in QPO properties are related to the innermost stable circular orbit.
A mass estimate of 1.9-2.1 ~$M_{\odot}$ has been given for 4U 1636-536
(\citealt{bom2005}).
Estimates for several systems 
give also very high mass values
with rather large uncertainties:
1.8--2.4~$M_{\odot}$ for Vela X-1\footnote{This range is based on the two
estimates given in \cite{q2003}: $1.88\pm 0.13$ and $2.27\pm 0.17 \,
M_{\odot}$.}
(\citealt{q2003}),
2.4$\pm 0.27$~$M_{\odot}$ for 4U 1700-37
(\citealt{c2002}; see also \citealt{h2003};
\citealt{vk2004}). \cite{s2004}
presented observations of a low-mass X-ray binary
2S~0921-630/V395~Car
for which the
1-$\sigma$ mass range for the compact object is 2.0--4.3~$M_{\odot}$.

 The existence of NSs with
$M\sim 1.8$--$2.4\, M_{\odot}$
is not in contradiction
with the present day theory of NS interiors. There are several models with
stiff equation of state  which allow the existence of NSs with masses
larger than
$2 ~ M_{\odot}$ (see a review and references in \citealt{ha2003, lp2005} 
and in
the contribution of J. Lattimer in these proceedings).
Here we will assume that masses of NSs with extreme rotation can reach the
maximum value $M_\mathrm{max}=3.45\, M_{\odot}$ according to the estimate by
\cite{ob1999}.  If, however, in nature $M_\mathrm{max}$ is lower than
this value, the part of the obtained mass spectrum exceeding it
should be attributed to black holes.

\subsection{Model}

 The model we use is discussed in detail in \cite{pp05}.
For our calculations we use the `Scenario Machine'' code developed
at the Sternberg Astronomical Institute.
\footnote{http://xray.sai.msu.ru/sciwork/scenario.html and
http://xray.sai.msu.ru/~$\sim$mystery/articles/review/.}
A description of most of the parameters of the code can be found
in \cite{lpp1996}.
The main assumptions of the scenario are:

\begin{itemize}
\item All NSs are born with $M=1.4\, M_{\odot}$.
\item At the common envelope stage a hypercritical accretion
(with $\dot M$ much larger than the Eddington value) is possible.
\item During accretion the magnetic field of a NS decays to a value
that cannot prevent rapid (maximum) rotation of the NS.
\item The Oppenheimer-Volkoff mass
of a rapidly rotating NS
(the critical mass of a BH formation)
is assumed to be 3.45~$M_{\odot}$ according to \cite{o2004}.
\end{itemize}

\begin{figure}[h]
\centering
  \includegraphics[width=250pt]{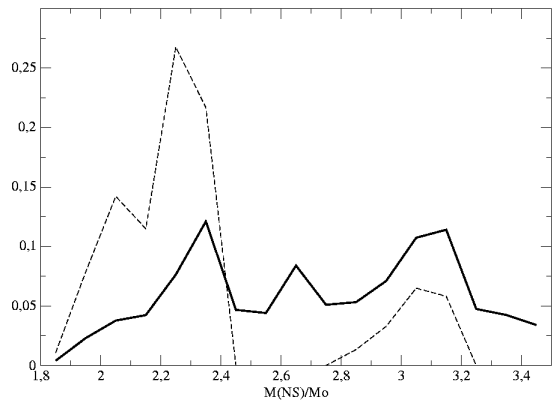}
\caption{Mass distribution of NSs. As we are interested only in the massive
population  we do not show the results for compact
objects with $M<1.8\, M_{\odot}$. The upper mass limit corresponds to SkyS
 with maximum rotation (Ouyed 2004).
The dashed line represents results for the scenario with zero
kick. The solid line is the non-zero kick. Left peaks for both distributions
correspond to NSs with a single episode of accretion. Right peaks are formed
by NSs which also increased their masses via accretion from WDs.
Distributions were normalized to unity, ie. the area below each line is
equal
to one.}
\label{fig:1}       
\end{figure}

\subsection{Mass spectrum}

As we are interested here in
systems with a high mass ratio 
it is necessary to consider three different
situations after NS formation when the secondary fills its Roche lobe:
{\it i.)} a normal star can fill its Roche lobe without common envelope
formation; {\it ii.)} a normal star can fill its Roche lobe with common
envelope  formation; {\it iii.)} a WD fills its Roche lobe.

To fill the Roche lobe a normal secondary
star has to evolve further than the main sequence stage.
During its evolution prior to the Roche lobe overflow
the mass of the star is nearly constant.
A common envelope is not formed if the normal
star is not significantly heavier than the NS. In this regime, mass is not
lost from the binary system. For more massive secondaries, formation of a
common envelope is inevitable, mass transfer is unstable. In this regime,
a significant fraction of the mass flow is lost from the system, so the mass
of the NS grows less effectively.
After the common envelope stage the
orbital separation becomes smaller, so later
on even a degenerate core of the secondary -- a WD -- can fill the Roche
lobe.

In the figure
we present our calculations of the mass spectrum of massive NSs.
About 25\% of accreting massive NSs have normal stars
as secondaries, the rest  have WD companions.
The formation rate of massive NSs was found to be $6.7 \,
10^{-7}$~yrs$^{-1}$.
This corresponds to $\sim10\,000$ of these compact stars in the Galaxy.


\section{Mass constraint}

The discussion below is related to our recent study of criteria to test
theoretical cooling curves \citep{petal06} based on an approach for hybrid 
stars with color superconducting quark matter cores \citep{gbv04}.

 Cooling curves of NS are strongly mass dependent (see contributions
by D. Page, A. Kaminker and others in this volume).
Unfortunately, a mass determination with high precision is available only
for NSs in binary systems.
Compact objects in X-ray binaries could accrete a significant amount of
matter. 
For some of the radio pulsars observed in  binaries, accretion also played
an
important r\^ole. Without any doubts masses of millisecond  pulsars do not
represent their initial values. However, there is a small number of NSs with
well determined masses, for which it is highly possible that these masses
did not change significantly since these NSs were born
(data on NS masses can be found, for example, in
\cite{Lorimer:2005bw} 
and references therein).
These are secondary (younger) components of double NS systems.
According to standard evolutionary scenarios these compact
objects never accreted a significant amount of mass. 
Their masses lie in the narrow range
1.18~-~1.39~$M_{\odot}$. 
Now there are nine double-NS
systems with well-estimated masses of the secondary components.
This set of data is a very good evidence in
favour of the mass spectrum used in our population synthesis calculations
\citep{petal06a}. 
Of course, some effects of
binary evolution can be important, 
and so for isolated
stars (or stars in very wide binaries)
the situation can be slightly different.
However, with these observational
estimates of initial masses of NSs we feel
more confident using the spectrum with a small number of NSs with
$M \stackrel{>}{\sim} 1.4$~-~$1.5 \, M_{\odot}$.

Brighter sources are easier to discover. So, among known cooling NSs the
fraction of NSs with {\it typical} masses, i.e. in the range
$1.1 \, M_{\odot} \stackrel{<}{\sim} M \stackrel{<}{\sim}  1.5 \, M_{\odot}$,
should be even higher than in the original mass spectrum.
So, we have the impression that it is
necessary to try to explain even cold (may be with an exception of 1-2
coldest) sources with $M\stackrel{<}{\sim}  1.4$~-~$1.5 \, M_{\odot}$.
Especially, the Magnificent seven and other young close-by
compact objects should be
explained as most typical representatives of the whole NS
population.
We want to underline that, even being selected by their observability in
soft X-rays, these sources form one of the most uniform samples of young
isolated NSs.
In this sense, the situation  where a significant
number of sources are explained by cooling curves corresponding to
$1.5 \, M_{\odot} \stackrel{<}{\sim} M \stackrel{<}{\sim} 1.7\, M_{\odot}$
should be considered as a disadvantage of the model.
Particularly, Vela, Geminga and RX J1856-3754 should not be
explained as massive NSs. 
  
   All the above gives us the opportunity to formulate the conjecture of a
{\it mass spectrum constraint}: data points should be explained 
mostly by NSs with {\it typical} masses.



\begin{acknowledgements}
SBP wants to thank the Organizers for support and hospitality.
The work of SBP was supported
by the RFBR grant 06-02-16025 and by the
the ``Dynasty'' Foundation (Russia).
The work of MEP -- by the  RFBR grant 04-02-16720 and that of 
HG by DFG grant 436 ARM 17/4/05.
\end{acknowledgements}




\end{document}